# RF Couplers for Normal-Conducting Photoinjector of High-Power CW FEL

Sergey Kurennoy, Dale Schrage, Richard Wood
*Los Alamos National Laboratory, Los Alamos, NM*
Tom Schultheiss, John Rathke
*Advanced Energy Systems, Medford, NY*
Lloyd Young
*TechSource, Santa FE, NM*

## Abstract

A high-current emittance-compensated RF photoinjector is a key enabling technology for a high-power CW FEL. A preliminary design of a normal-conducting, 2.5-cell $\pi$-mode, 700-MHz CW RF photoinjector that will be built for demonstration purposes, is completed. This photoinjector will be capable of accelerating a 100-mA electron beam (3 nC per bunch at 35 MHz bunch repetition rate) to 2.7 MeV while providing an emittance below 7 mm-mrad at the wiggler. More than 1 MW of RF power will be fed into the photoinjector cavity through two ridge-loaded tapered waveguides. The waveguides are coupled to the cavity by "dog-bone" irises cut in a thick wall. Due to CW operation of the photoinjector, the cooling of the coupler irises is a rather challenging thermal management project. This paper presents results of a detailed electromagnetic modeling of the coupler-cavity system, which has been performed to select the coupler design that minimizes the iris heating due to RF power loss in its walls.

This note describes the normal-conducting RF cavity of the high-current CW photoinjector for high-power FELs. We concentrate on three-dimensional (3-D) calculations for the RF input couplers and ridge-loaded tapered waveguides. In particular, the method of calculating the RF coupler irises and power distribution on them is described in detail. It is based on 3-D time-domain calculations with the CST MicroWave Studio.

# 1. Introduction.

The normal-conducting RF cavity of the high-current CW photoinjector (PI) for high-power free-electron lasers consists of 2.5 cells plus a vacuum plenum, with the on-axis electric coupling through large beam apertures, and has an emittance-compensating solenoid with a bucking coil. The cavity design is illustrated in Fig. 1, and described in publications [1]. After the first half-length cell (on the left), where a photocathode will be housed, there are two full-length ones, followed by a vacuum plenum with eight vacuum-pump ports attached. Figure 1 shows cooling water pipes, as well as two ridge-loaded tapered waveguides for RF input, which are connected to the third cell. The vacuum-plenum cell has its resonance frequency at about 650 MHz, well below the frequency 700 MHz of the working π-mode in the cavity.

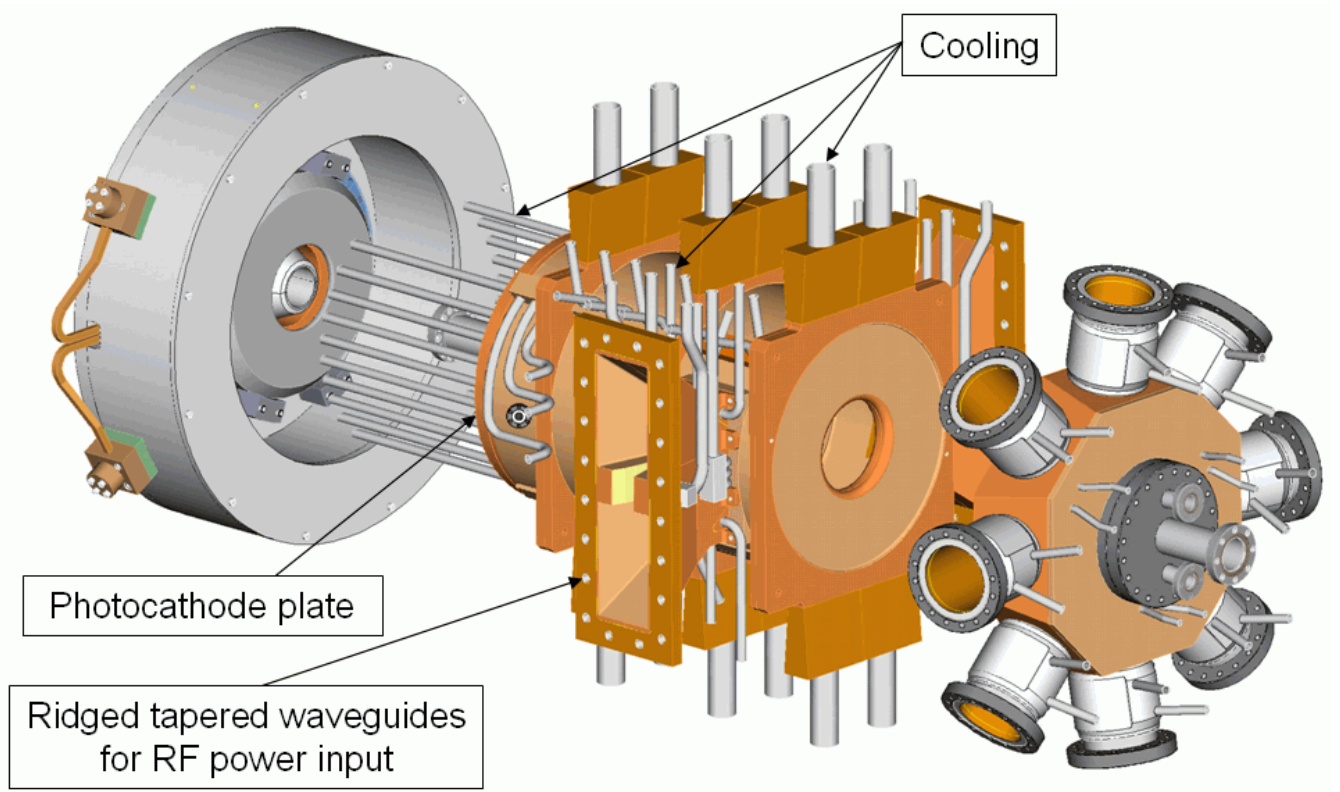

Fig. 1: Photoinjector RF cavity with vacuum plenum (right) and emittance-compensating magnets (left).

The basic 2-D design of the photoinjector RF cavity was performed by Lloyd Young using the Poisson / Superfish (SF) codes [2] and Parmela beam dynamics simulations. Figure 2 shows the electric field lines as calculated by SF for the cavity. This is the most recent, so-called 7-7-5, design of the PI cavity, for which the electric field gradients are 7 MV/m in the first two cells and 5 MV/m in the third one.

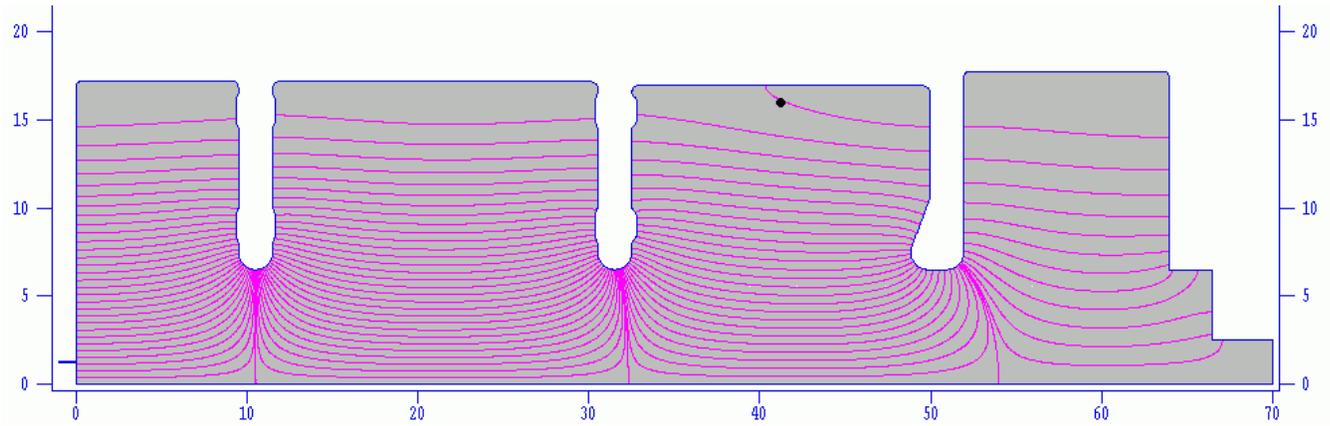

Fig. 2: SF output for 2.5-cell PI RF cavity (dimensions in cm): electric field lines.

One can see tuning rings on the separators, i.e. on the cavity walls separating cells 1 & 2, and 2 & 3. The profile of the longitudinal electric field on the cavity axis is shown in Fig. 3 for the 7-7-5 cavity design. The electric field alternates its direction from one cell to the next, as it should be in a π-mode. Compared to the old, 7-7-7 design, where the electric field gradient was 7 MV/m in all three cells, here the fields in the third cell are reduced. This change does not have an adverse effect on the beam emittance, because the beam energy is already high enough after the first two cells, but it allows reducing the power density on the coupler irises in the third cell, as will be shown below. In both cases, the fields in the vacuum plenum are low.

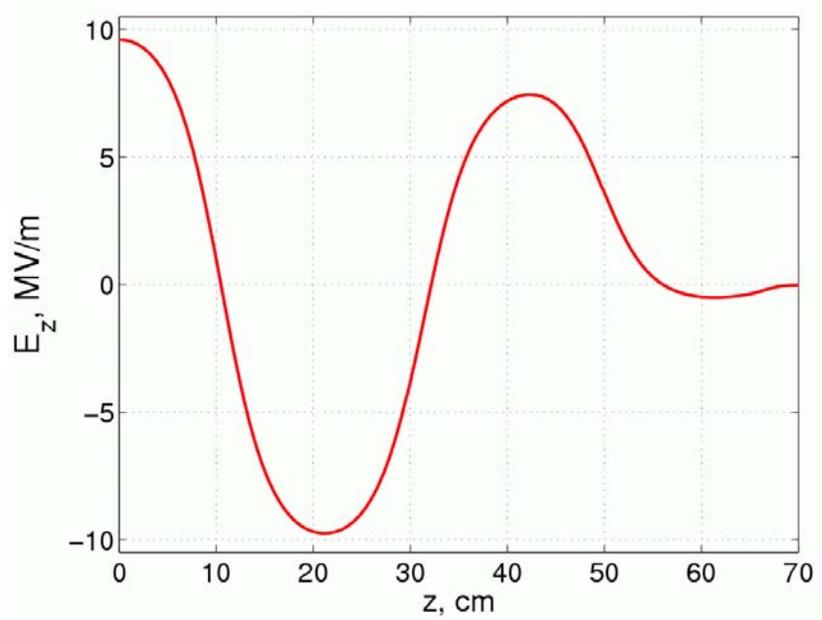

Fig. 3: On-axis electric field profile for 7-7-5 design (from Microwave Studio eigensolver).

The frequency shifts due to 3-D details of the photoinjector RF cavity (vacuum pump ports, coupler irises) have been calculated with MicroWave Studio (MWS) [3] eigensolver. The calculations are described in tech note [4]. These frequency shifts have been taken into account in the Superfish design, so that the frequency of the working mode will be at 700 MHz when the effects of the 3-D cavity details are included.

Some relevant parameters of the photoinjector are summarized below in Table 1. While this design is for the electron beam current of 100 mA (3 nC per bunch at 33.3 MHz bunch repetition rate), the current in the photoinjector cavity can be increased above 1 A by increasing the bunch repetition rate. More information on the PI design and parameters can be found in the presentations at the design review [5].

Table 1: Parameters of the high-current CW FEL photoinjector.

| Parameter | Value | Comment |
|---|---|---|
| Exit beam energy, MeV | 2.54 | |
| Beam current, mA | 100 | |
| Total beam power, kW | 254 | |
| Wall power loss, kW | 668 | at 20°C, from SF calculations |
| Wall power loss from thermal analysis, kW | 728 | resistivity at surface temperature |
| Beam transverse rms emittance, mm·mrad | < 7 | after a drift, at 90 cm from PI |
| Max wall loss power density, W/cm$^2$ | 103 | at 20°C, from SF calculations |
| Max wall loss power density, W/cm$^2$ | 114 | thermal analysis, resist. at surface temp. |

The surface current distribution in the PI cavity for the 7-7-5 design, as calculated by MWS eigensolver, is plotted in Fig. 4. The photocathode position is (0, 0, 0) in the coordinates shown in Fig. 4. As one can see, the reduced gradient in the third cell leads to significantly lower values of the wall power loss density in that cell, compared to the first two cells. It is about 43 W/cm$^2$ at the location where the RF coupler iris will be attached. For comparison, the same density for the 7-7-7 design was 75 W/cm$^2$.

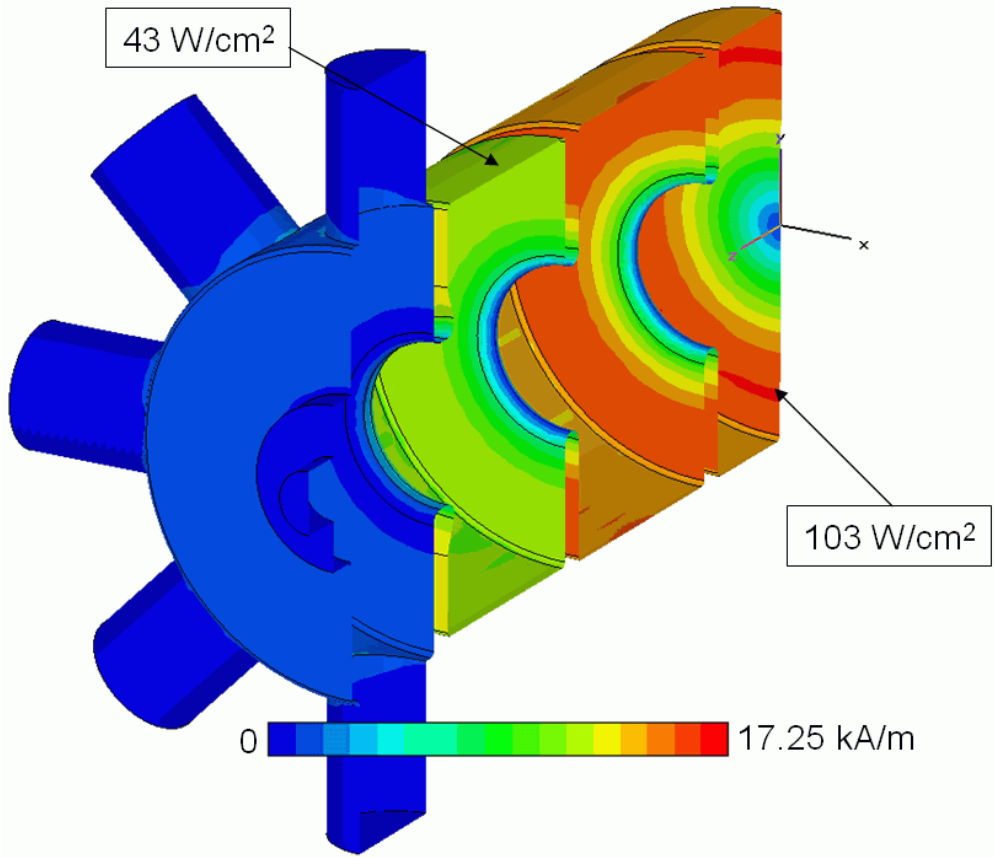

Fig. 4: Surface current distribution in the photoinjector cavity for 7-7-5 design.

## 2. Ridge-loaded tapered waveguides.

As one can see from Table 1, almost 1 MW of RF power is needed for CW operation of the photo-injector cavity. The RF power will be fed into the cavity through two ridge-loaded tapered waveguides (RLWG), as shown in the schematic of Fig. 1. Figure 1 shows only the tapered ridge-loaded section of the RF input waveguide, with its larger cross section equal to that of a half-height waveguide WR1500. This section will be connected to another transition, from the half-height to a standard full-height WR1500, as shown in Fig. 5.

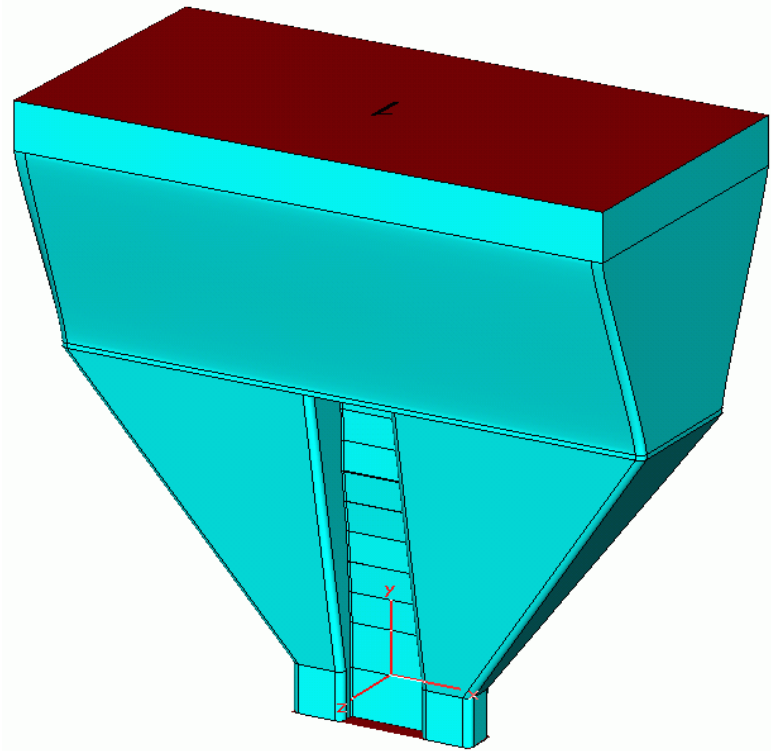

Fig. 5: MWS model of transition from full-height WR1500 (top) to half-height WR1500, followed by ridge-loaded tapered waveguide (RLWG).

The length of the tapered RLWG, about 16 cm, was limited by manufacturing restrictions: the PI cavity and two tapered RLWG have to fit into a brazing furnace. The design of RLWG is based on experience from the LEDA RFQ and SNS power couplers. In this case, the matching ridge was made narrow to minimize a possibility of multipacting. The ridge profile was designed [6] using 2-D SF calculations for multiple cross sections corresponding to the horizontal (parallel to $x$) lines in Fig. 5. The profile can be seen in detail in Fig. 6, as well as in the next Section. The ridge purpose is to minimize reflections of input RF power due to waveguide tapering.

3-D S-parameter computations with the MWS for the model shown in Fig. 5 have confirmed that the input signal reflections in RLWG are indeed small at 700 MHz.

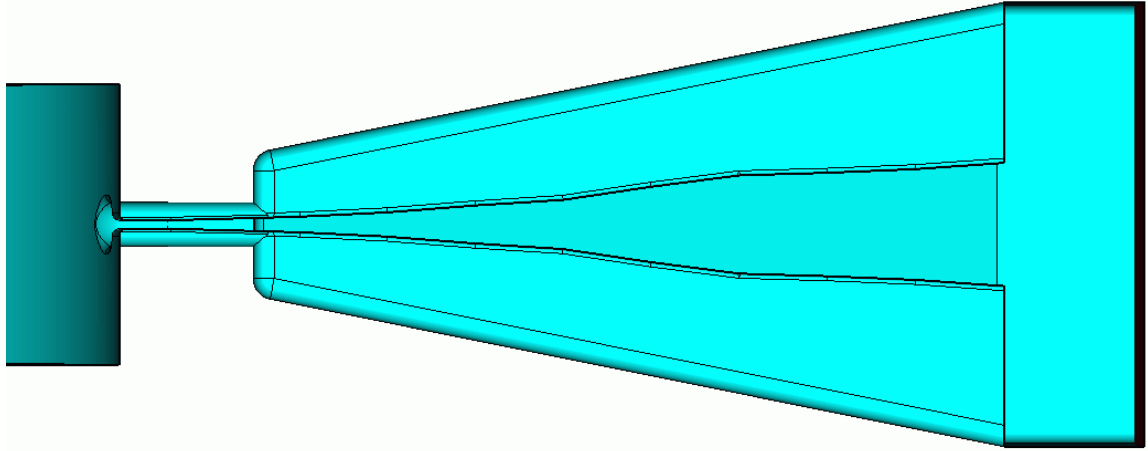

Fig. 6: Ridge profile in the ridge-loaded tapered waveguide (MWS model).

## 3. RF couplers for FEL photoinjector cavity.

The ridge-loaded tapered waveguides are connected to the third cell of the photoinjector cavity via "dog-bone" shaped coupler irises in the thick cavity wall. Figure 7 shows details of the coupler irises. One should note that all pictures here show only the vacuum (inner) parts of the cavity; these parts are supposed to be surrounded by metal walls. The wall separating the PI cavity from the tapered RLWG is about 1.2″ thick. The iris consists of a 2″-long narrow slot with two cylindrical holes near its ends; the outer ends of the holes are blended inside the cavity to reduce the enhancement of the cavity magnetic field near these ends. The coupling is adjusted by changing the diameter of the holes.

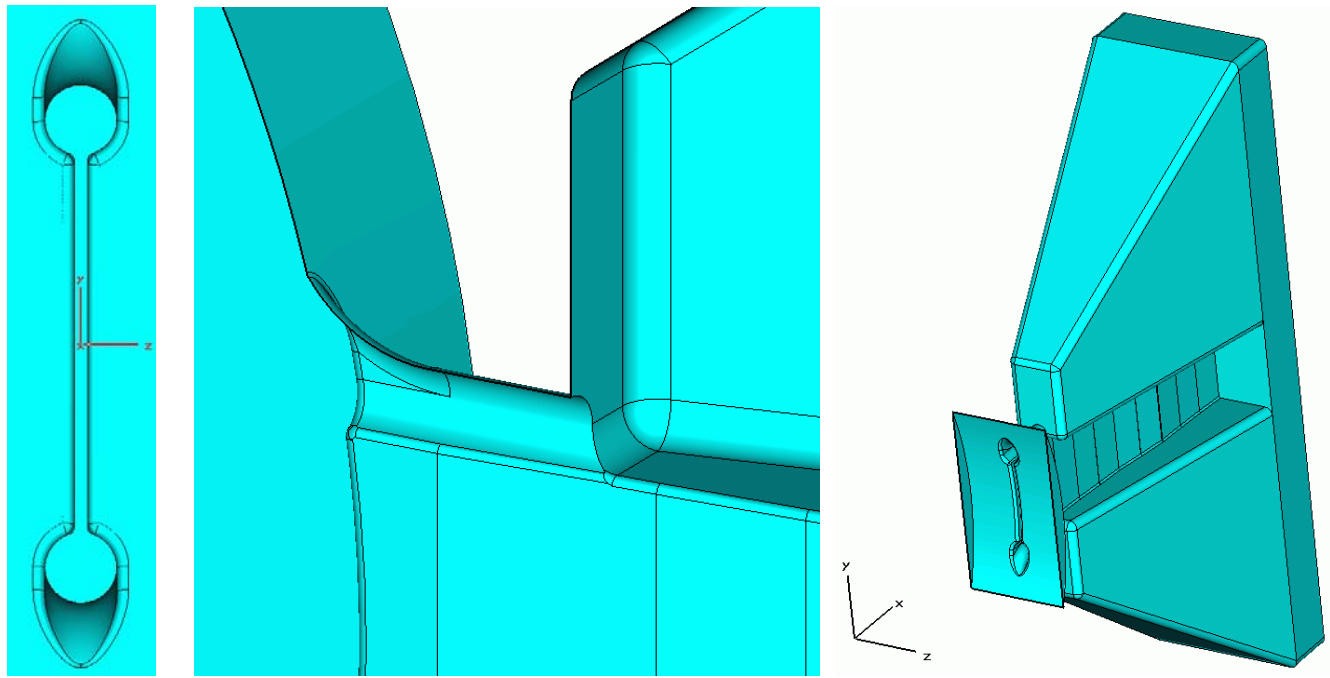

Fig. 7: Coupler irises: a view from inside the PI cavity – left, a cut-out view through the middle of the iris gap ($z = 0$ plane) – center, and a view with the ridge-loaded tapered waveguide – right.

For calculating the coupling we use a MWS model consisting of a pill-box cavity with two attached tapered ridge-loaded waveguides, as shown in Fig. 8, instead of simulating the whole PI cavity. The pill-box cavity is essentially a slice cut out of the third cell of the PI cavity. It has the same radius as the third cell, 169.24 mm. The frequency of the $TM_{010}$-like mode of this pill-box without the couplers is adjusted to be 700 MHz by choosing the axial length of the on-axis cylindrical extension. To make sure that the couplers can be manufactured exactly as calculated, the model layout was created by a CAD system (Pro/E) at AES [7], and then imported into the MWS.

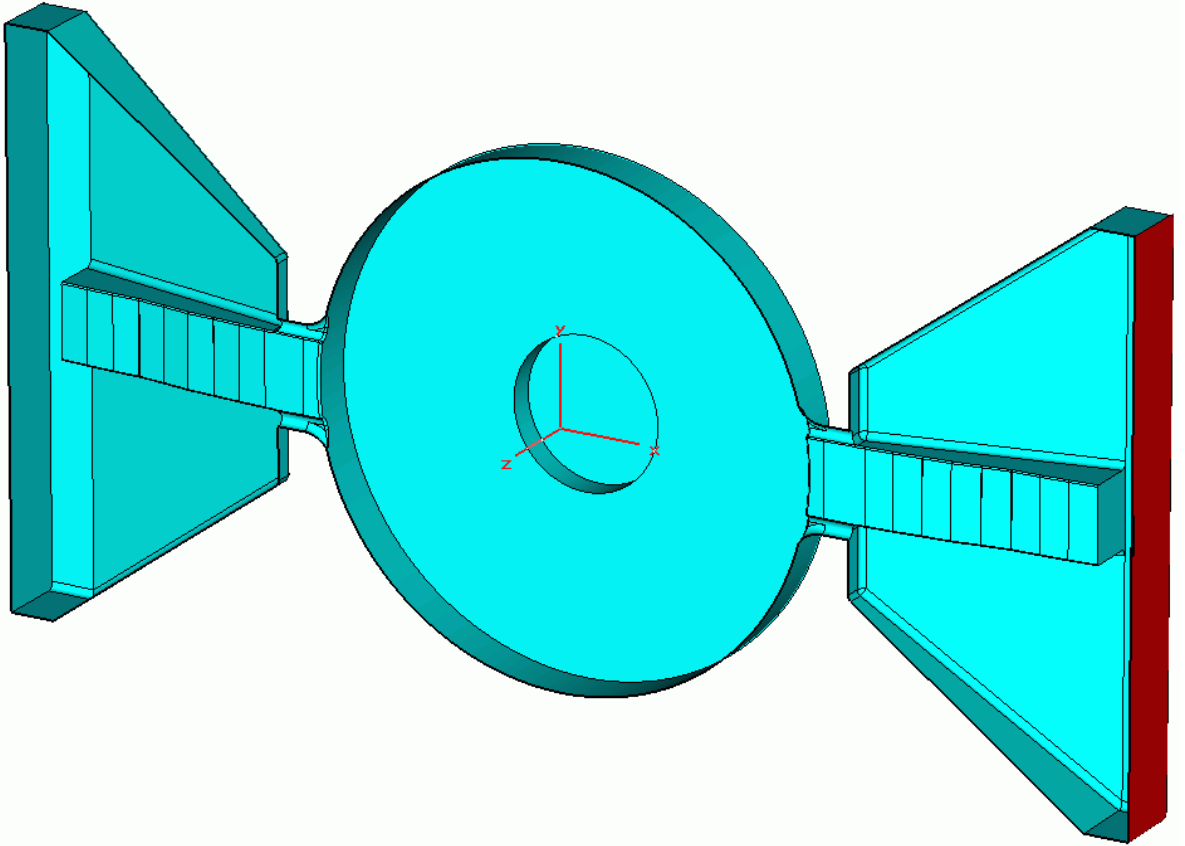

Fig. 8: MWS model of RF couplers: a cut-out view through the iris mid-plane ($z = 0$) – cf. Fig. 7.

The required waveguide-cavity coupling for the 2.5-cell PI cavity with the beam current 100 mA can be found using values from Table 1:

$$\beta_c = \frac{P_w + P_b}{P_w} = 1.38, \tag{3.1}$$

where $P_w$ is the wall power loss and $P_b$ is the beam power. Of course, when the same couplers are connected to the pill-box cavity in our model, the coupling will be different. It is related to the coupling (3.1) as

$$\beta_{pb} = \beta_c \frac{W_c}{W_{pb}} \left( \frac{H_{pb}}{H_c} \right)^2 \frac{Q_{pb}}{Q_c}, \tag{3.2}$$

where $W_i$, $H_i$, $Q_i$ are the stored energy, magnetic field at the coupler location (without coupler), and unloaded quality factor for the cavity with index $i = c, pb$. All quantities entering Eq. (3.2) can be easily found with eigensolvers. From Eqs. (3.1-2), one can find the required value of the external quality factor for the pill-box model:

$$Q_e = \frac{Q_c}{\beta_c} \frac{W_{pb}}{W_c} \left(\frac{H_c}{H_{pb}}\right)^2 = 1933. \tag{3.3}$$

We use direct MWS time-domain calculations to calculate the external quality factor in the model, and to adjust the coupling in such a way that it matches the value in (3.3). This technique was described in Ref. [8] and consists of the following. From the problem symmetry, we use in MWS calculations only 1/8 of the geometry in Fig. 8, imposing magnetic boundaries at $x = 0$ and $y = 0$, and an electric one at $z = 0$. First, the structure is fed with a short RF pulse through the waveguide port shown in red in Fig. 8. The pulse excites the fields in the pill-box cavity, which then decay due to radiation through the coupler irises into the waveguides, cf. Fig. 9. This radiation is the only sources of energy loss, since all metal surfaces are assumed perfectly conducting; therefore, the field decay constant is directly related to the model cavity external quality factor $Q_e$. Figure 9 plots the amplitude of the electric field in the center of the cavity versus time; the plot includes hundreds of the field oscillations.

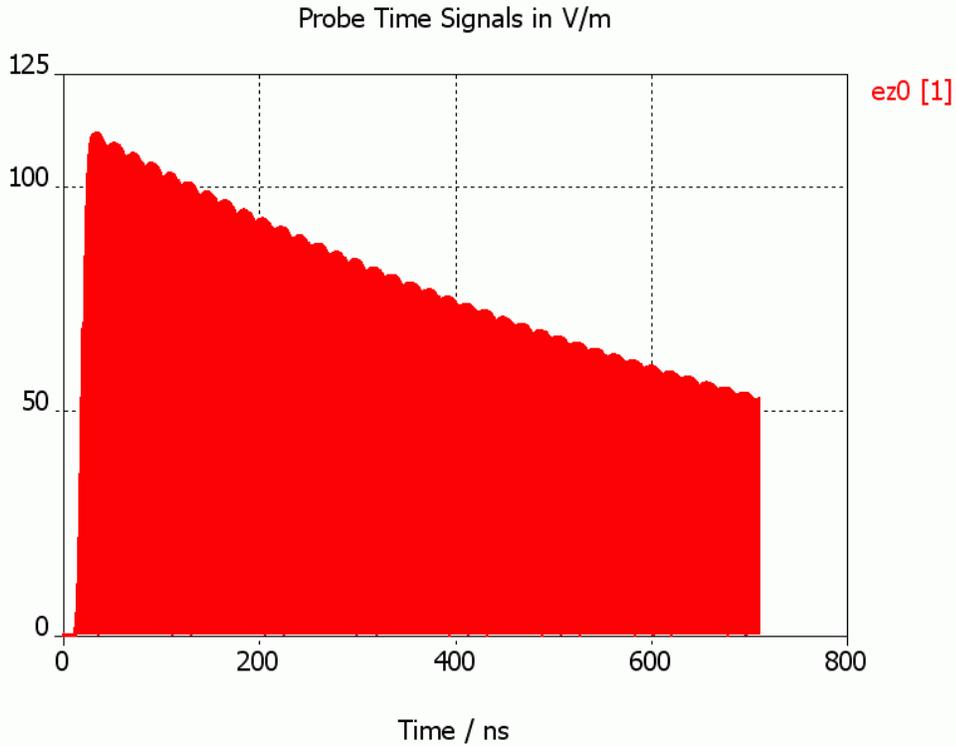

Fig. 9: Electric field decay from MWS time-domain calculations.

The dependence of the field energy on time is even more convenient for calculating the cavity external quality factor $Q_e$, as illustrated in Fig. 10. The slope of the exponential (linear on dB scale) part of this plot gives $Q_e$ directly:

$$Q_e = \frac{20\pi f(GHz)}{\ln 10} \frac{\Delta t(ns)}{\Delta E(dB)}. \tag{3.4}$$

For the particular case shown in Fig. 10, from Eq. (3.4) the external quality factor $Q_e = 2010$. Therefore, the coupling is a bit low, cf. (3.3), and the radius of the iris holes should be slightly increased to increase it. The coupling dependence of the hole radius $r$ is rather strong, faster than $r^3$. For the iris layout shown in Figs. 7-8, with the iris slot width 1.788 mm and the hole blend radius inside the cavity 19 mm, the correct coupling is achieved for the hole diameter equal to 9.5 mm.

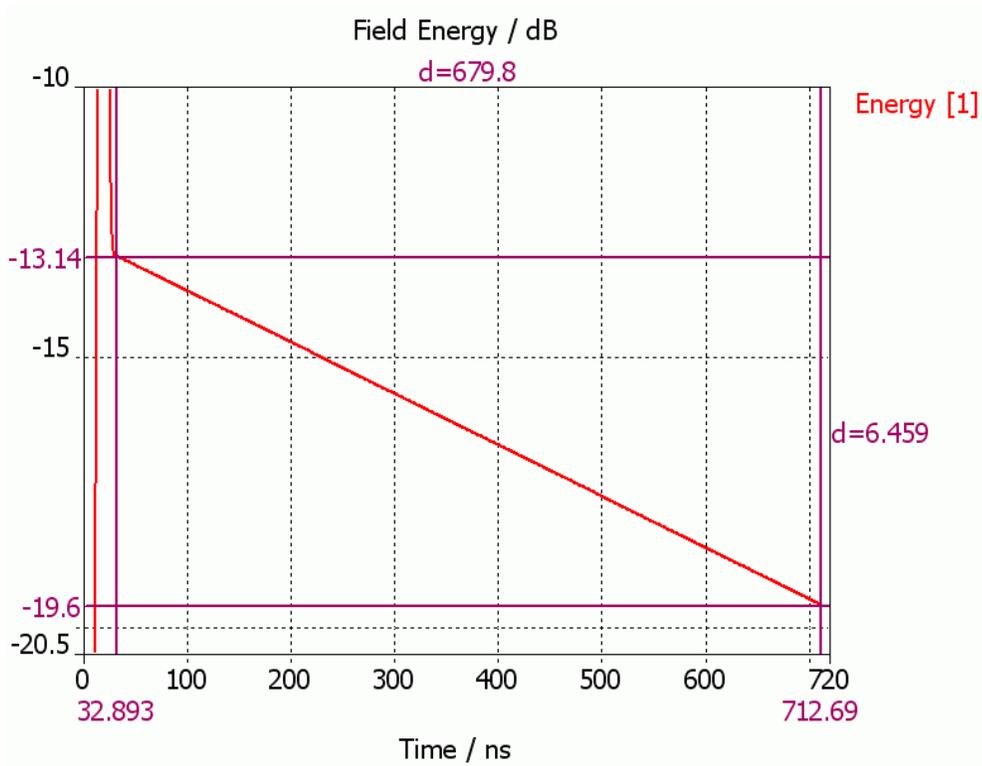

Fig. 10: Field energy decay (red) from time-domain calculations and plot measure lines (magenta).

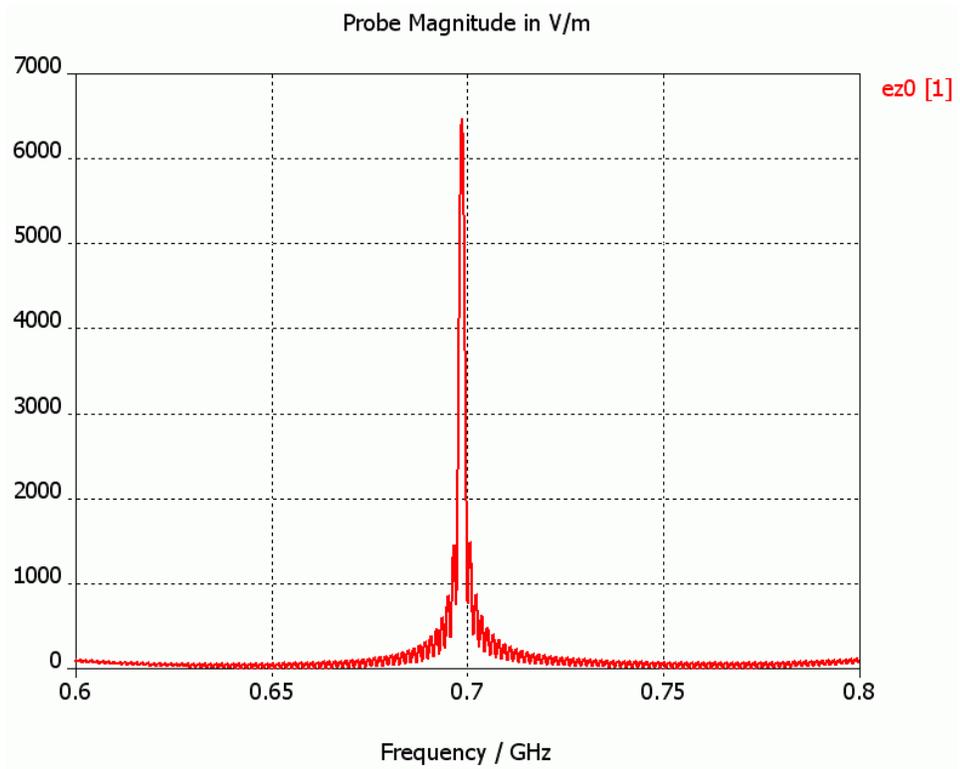

Fig. 11: Fourier-transform of electric field time signal in Fig. 9.

Fourier-transform of the time signal in Fig. 9 gives the resonant frequency $f_{res}$ of the model with RLWGs, as shown in Fig. 11. As one can expect, this frequency is lower than the frequency of the pill-box cavity without RLWGs, due to the magnetic field penetration into the waveguides. In this particular case $f_{res}$ = 698.727 MHz.

After the coupler iris parameters are adjusted to give the correct coupling, another MWS time-domain simulation is performed to find the field distributions in the model. This time the RF input signal has its frequency equal to $f_{res}$, found earlier from calculations with a short Gaussian RF pulse, and its amplitude gradually increases during 100 ns, and after that remains constant at a certain level. Figure 12 shows both the input and output signals at the waveguide port versus time. Again, as in Fig. 9, these plots include hundreds of the field oscillation periods; we see only signal envelopes. While the waveguide input remains constant, the output decreases reaching a point where it vanishes, and increases again after that. The output decrease is due to a destructive interference of two waves: one is reflected from the coupler iris, and the other is radiated into the waveguides from inside the cavity. The reflected-wave amplitude remains constant when the input is constant, while the radiated-wave amplitude increases as the cavity field increases. These two waves are always in opposite phases [9]. As a result, at certain moment the two waves cancel each other, so that the reflected power vanishes at that particular moment. This situation corresponds to an exact match; therefore, the field snapshots at that moment give us field distributions for the matched (100 mA) case.

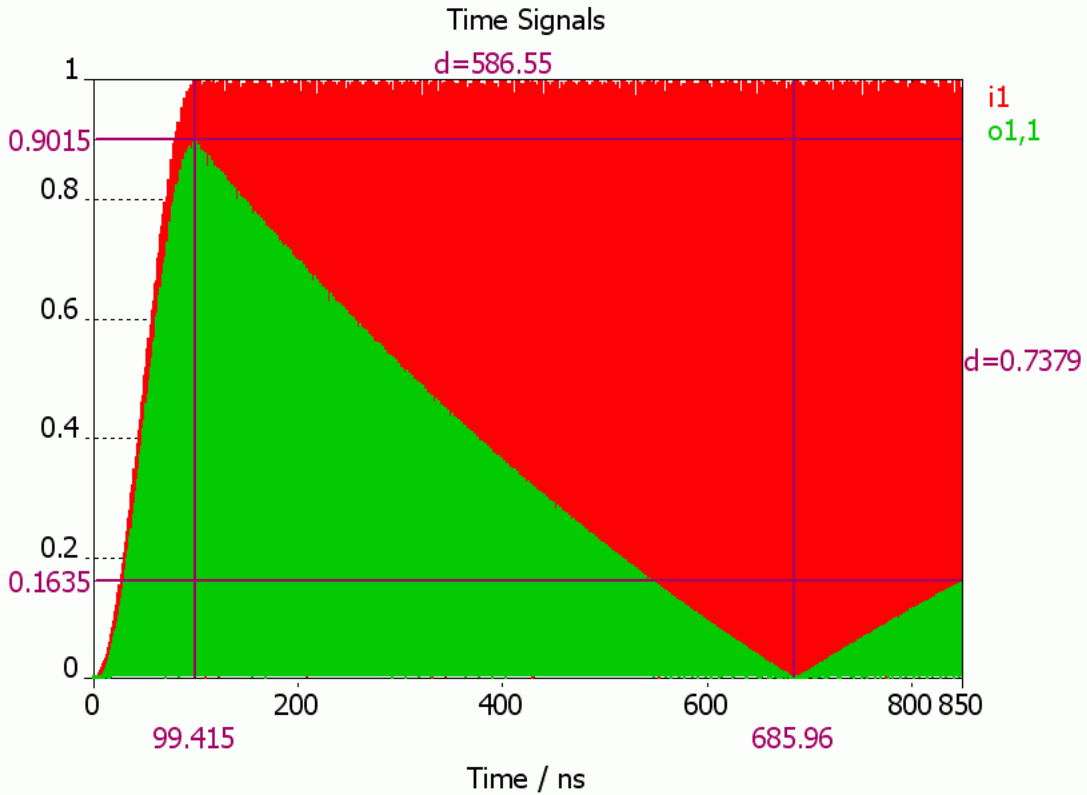

Fig. 12: Waveguide input (red) and output (green) signals from MWS time-domain calculations.

Another interesting point marked on the output signal plot in Fig. 12 is where its amplitude reaches about 16% of the RF input amplitude. This point corresponds to the thermal-test situation, when the photoinjector cavity will be tested running with the nominal field gradients but without beam. One can calculate that in that case, due to a mismatch, the power reflected back into the waveguides is equal to 2.5% of the input RF power, as explained below.

If we assume that the fields in the photoinjector cavity are brought to their nominal values, the power dissipated in the cavity walls is fixed and equal to $P_w$, cf. Table 1. The energy balance can be written as

$$P_{in} - P_{out} = P_w + P_b \equiv \beta P_w, \qquad (3.5)$$

where $P_{in}$ and $P_{out}$ are the RF input power and the total power reflected back into the waveguides. Here we introduced an arbitrary coefficient $\beta$ as the ratio of the total power delivered to the cavity to the wall power loss. For the 100-mA matched case, when $\beta = \beta_c = 1.38$, there will be no reflected power: $P_{out} = 0$, cf. Eq. (3.1). In the case of a mismatch, i.e. for other values of $\beta$, in particular, when there is no beam in the cavity ($\beta = 1$), there will be some power reflection. Using the physical picture discussed above, that there are two waves propagating back in the waveguides – one reflected from the coupler iris and another radiated through the iris by the cavity fields, – which always have opposite phases, one can calculate the power reflection analytically. The ratio of the total reflected power to the RF input power is given by the following expression:

$$\frac{P_{out}}{P_{in}} = y \left[ \frac{x - \sqrt{y + (1-y)x}}{y + x} \right]^2. \qquad (3.6)$$

It depends on two dimensionless parameters: $x = \beta/\beta_c$, and $y$, which is equal to the ratio of the power reflected by the coupler iris to the input power. We know that $y$ should be rather close to unity, $1 - y \ll 1$, because initially almost all power is reflected back to the waveguide from the iris, cf. Fig. 12. In fact, dependence on $y$ in (3.6) is rather weak for values of $x$ close to 1, as one can see from Fig. 13, which plots the ratio (3.6) for a few different values of $y$. The black dot in Fig. 13 corresponds to the thermal-test point with $x = 1/\beta_c = 0.725$; the power ratio is about 0.025 at that point. One should emphasize again that we used the assumption of the fixed (nominal) fields in the cavity, i.e. the fixed power loss $P_w$, in deriving Eq. (3.6).

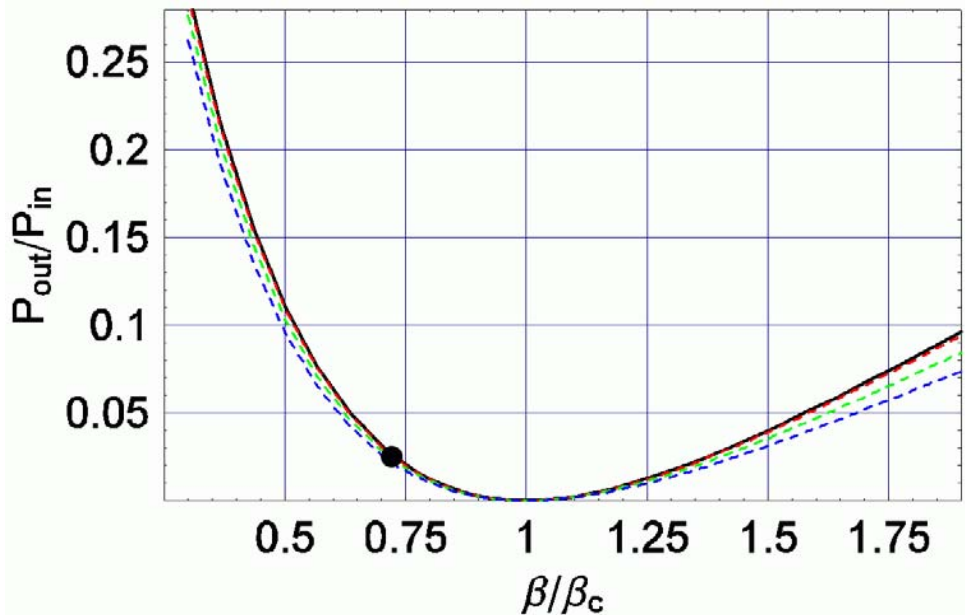

Fig. 13: Ratio of the reflected to the input power versus mismatch parameter $x = \beta/\beta_c$ for $y = 1$ (black solid curve), 0.98, 0.9, and 0.81 (red, green, and blue dashed curves).

Figures 14-15 show the snapshots of the magnetic fields and surface currents taken at the match point.

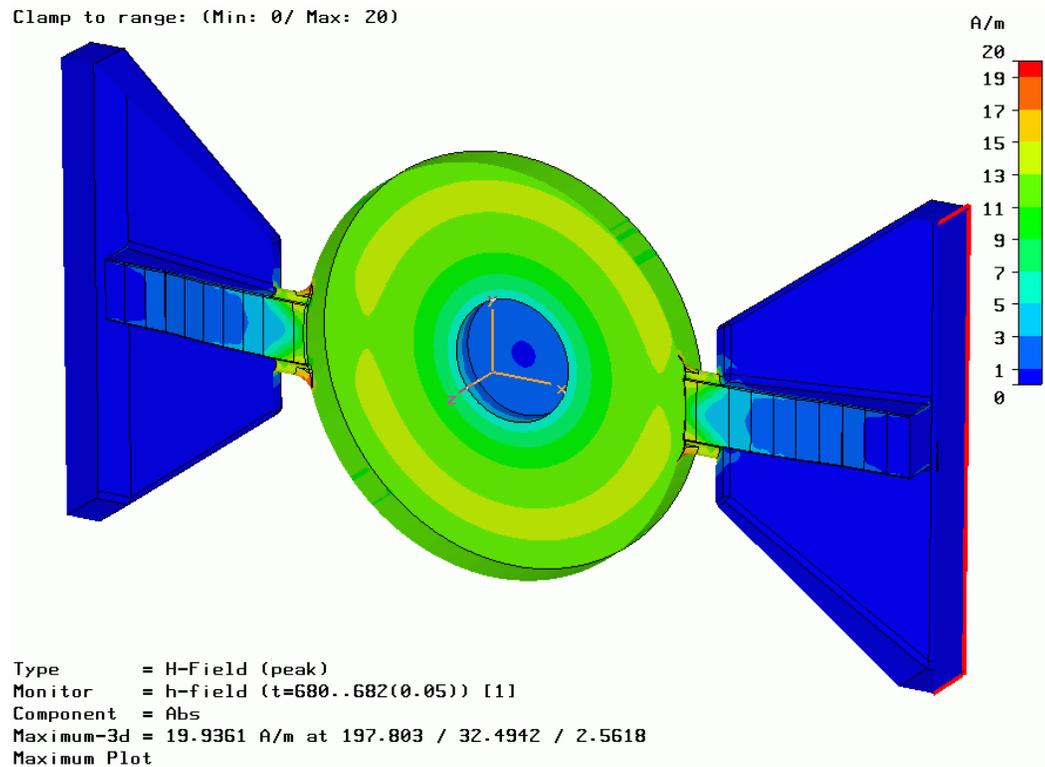

Fig. 14: Surface magnetic fields at the match point from MWS time-domain calculations.

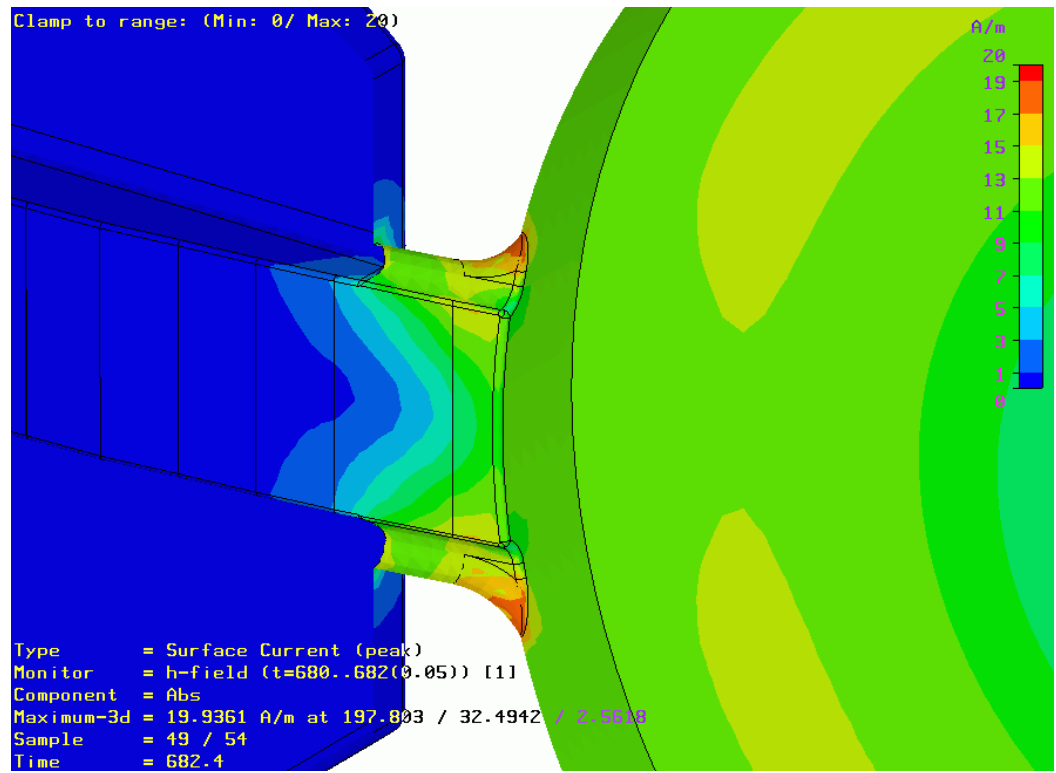

Fig. 15: Surface currents near the coupler iris at the match point from MWS time-domain calculations.

One can see that the maximal field values are near the ends of the coupler irises. One should note that the field scaling in Figs. 14-15 corresponds to the MWS default for time-domain simulations: 1 W peak power (0.5 W average power) through a waveguide port. Since for the matched case of 100-mA beam we need the total RF power of 922 kW, cf. Table 1, our RF input should be 461 kW per waveguide. This gives us the field scaling factor of $\sqrt{922000} = 960$ for the fields at the match point, Figs. 14-15. As the result, the max power density near the iris ends is about 120 W/cm$^2$, assuming the cavity surface conductivity $\sigma = 5.8 \cdot 10^7$ 1/(Ohm·m). The regions of high power density are small and well localized, which makes easier their cooling with dedicated cooling channels, see in [5].

A similar field snapshot for the thermal-test point is presented in Fig. 16. The field scale in this picture is different from that in Figs. 14-15, but the scaling factor is also different. At the thermal-test point, the total RF power is the sum of the wall loss plus 2.5% reflected power, which gives us $1.025 \cdot P_w = 684.7$ kW. The RF power input should be 342 kW per waveguide. Then the field scaling factor is $\sqrt{687400} = 827$ for the thermal-test point, which results in the same value for the max power density near the iris ends, about 120 W/cm$^2$, as for the 100-mA match point.

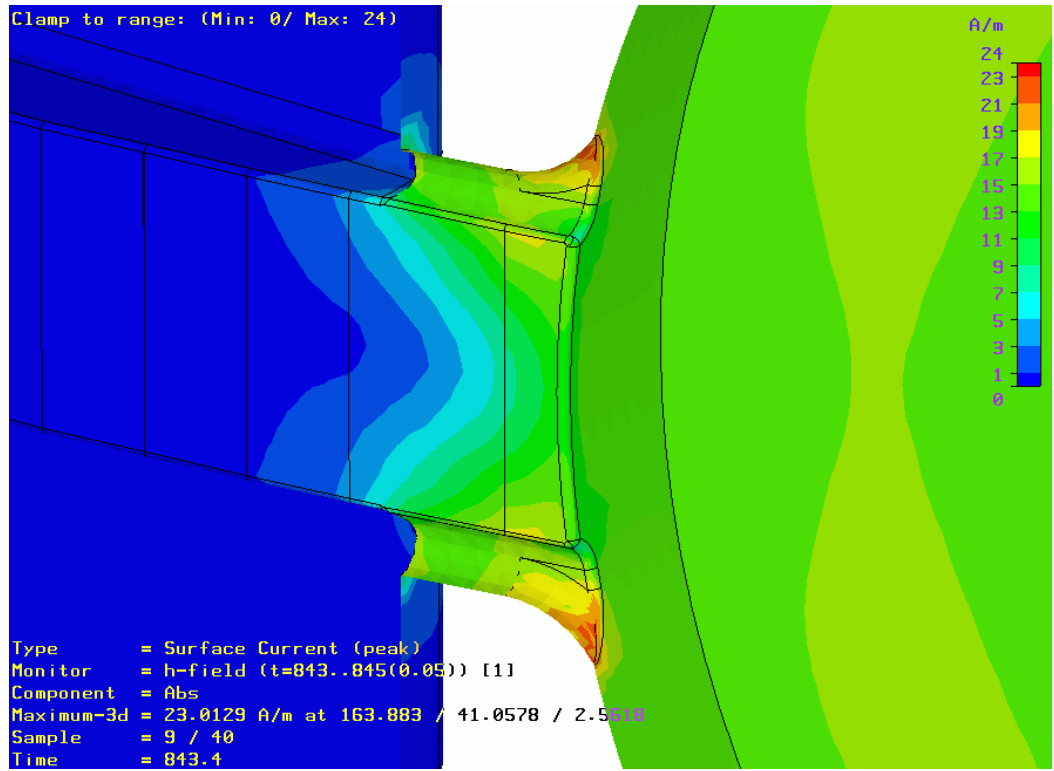

Fig. 16: Surface currents near coupler irises at the thermal-test point (no beam, 2.5% reflected power) from MWS time-domain calculations.

This result validates our plan that the thermal management of the cavity should be demonstrated first without beam, because it will be the same with the beam operation. It also shows that the hottest spots are defined by the fields inside the cavity, not by the amount of the RF power fed through waveguides. As one can see from Figs. 14-16, the power loss density in the tapered RLWG is relatively low.

Because the field structure inside the cavity is similar to that of the corresponding cavity eigenmode, one can use an eigensolver to cross-check the max field values near the coupler irises. To do that, we cut the RLWG at about $\lambda/8$ distance from the iris, where $\lambda = 52$ cm is the wavelength at 700 MHz in the RLWG, and impose an electric boundary condition in the cut plane. The exact cut distance is not very important because the fields in the RLWG are low compared to those in the cavity. Of course, such calculations do not give correct fields in waveguides. However, this eigenvalue problem can be solved much faster than time-domain runs above, and one can easily go to rather fine meshes. The results for the fields calculated with the MWS eigensolver are shown in Fig. 17. In this particular case, the mesh was 3.006 million mesh points for 1/8 of the geometry. The field scaling for eigenmode solutions can be found by comparison of the surface magnetic field values far from the irises with those in the photo-injector cavity at the same location. (In fact, the same scaling procedure can be used for the time-domain field snapshots; it gives results close to those obtained using the RF input power scaling as described above). The magnetic field of the eigenmode is equal to 17.6 kA/m far from the irises in the pill-box mid-plane cross section. For the nominal gradients in the 7-7-5 design, the field value for the same cross section in the third cell of the photoinjector cavity is 11.2 kA/m. Therefore, the scaling is 11.2/17.6 = 0.636 for the fields (surface currents) in Fig. 17. This gives the max power density near the iris ends 118 W/cm$^2$, which is close to the time-domain results.

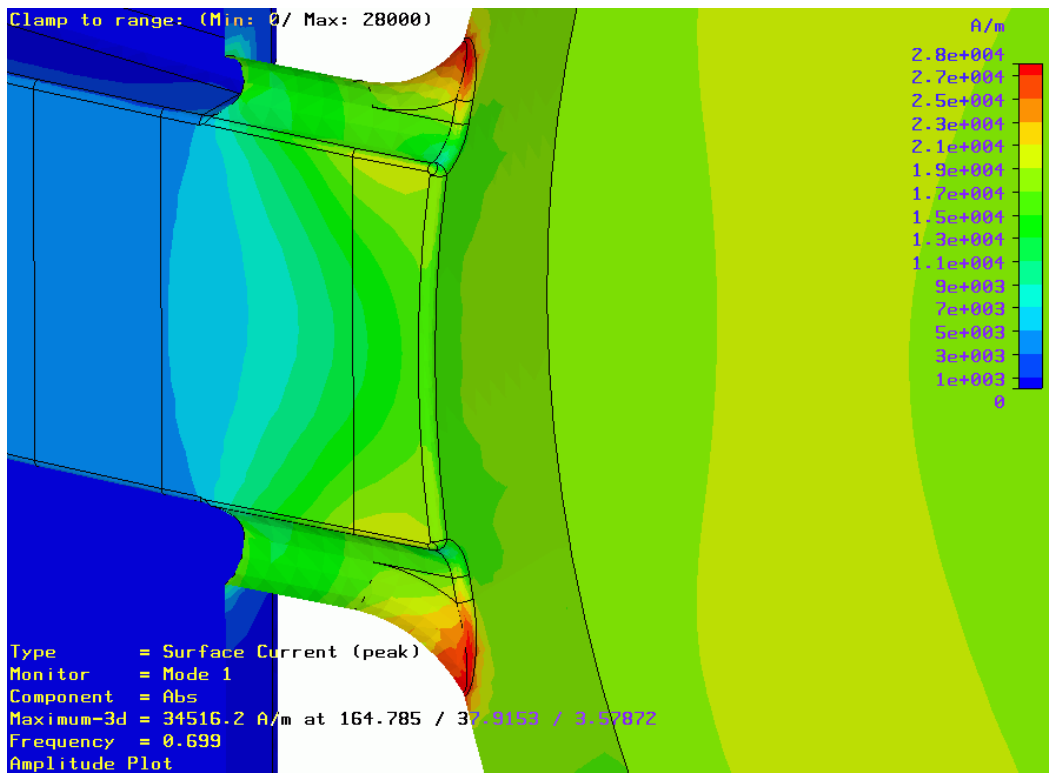

Fig. 17: Surface currents near the coupler irises from MWS eigensolver calculations.

Results from MWS time-domain and eigensolver computations of the maximal power density on the coupler irises for the 7-7-5 design of the photoinjector cavity are summarized in Tables 2-3. The results are given versus the number of mesh points for 1/8 of the geometry used in calculation, with account of the problem symmetry. Table 2 shows results for both the match point (100 mA beam current, 461 kW RF power input per waveguide) and the thermal-test point (no beam, 2.5% power reflection due to mismatch, 362 kW RF power input per waveguide), for the same mesh. One can see that the values for crude meshes are somewhat low; this is mainly because mesh cells are still too large to catch the localized hot spots near the iris ends. As mesh densities increase, the maximal power density values tend to stay around 120 W/cm$^2$. One should mention that the maximal power densities have been calculated also at AES [10] with an eigensolver in the finite-element code ANSYS that was used for the thermal analysis, and found to be in agreement with those from MWS calculations.

Table 2: Max power density from MWS time-domain computations.

| Mesh, K points | Max power density, W/cm$^2$ |
|---|---|
| 111 | 107 |
| 111* | 104* |
| 312 | 120 |
| 312* | 119* |
| 760 | 114 |
| 760* | 114* |
| 773 | 118 |

\* Thermal-test point

Table 3: Max power density from MWS eigensolver computations.

| Mesh, K points | Max power density, W/cm$^2$ |
|---|---|
| 86 | 95 |
| 201 | 109 |
| 734 | 120 |
| 773* | 121* |
| 1539 | 122 |
| 3006 | 118 |

\* For the layout of Fig. 8, w/o waveguide cuts.

Based on the results in Tables 2-3, we conclude that the maximal power density on the coupler irises for the 7-7-5 design is 120 W/cm$^2$. This value should be compared to the power density on the smooth walls in the cross section where the coupler irises are located, in the third cell of the photoinjector cavity, i.e. to 43 W/cm$^2$, cf. Fig. 3. The ratio of these two power densities is 2.79, which means that the magnetic field enhancement due to the iris presence is only by a factor of $\sqrt{2.79} = 1.67$. For comparison, in the LEDA RFQ couplers, such a power ratio was about 10, while the maximal power density near the iris ends was around 150 W/cm$^2$ [11]. In the presented design, the coupler irises create significantly smaller distortions of the cavity magnetic field, which allows us to keep the maximal power density below that in the LEDA RFQ couplers, even though the average power density on the smooth walls is almost three times higher than that in the LEDA RFQ.

Another useful comparison is with the case of the 7-7-7 design for the photoinjector cavity. In that case, the maximal power density on the coupler irises was calculated at 220 W/cm$^2$ [1], mainly because of the higher value of the smooth-wall power density, 75 W/cm$^2$. While the ratio of these two densities is still below 3, the maximal density is higher than what was already demonstrated in the LEDA RFQ CW operation. This fact was the main reason for changing the photoinjector cavity design to 7-7-5.

## 4. Feasibility of RF couplers for 1-A photoinjector cavity.

Here we present some estimates on possible RF couplers required for a photoinjector cavity designed for 1 A of the electron beam current. We assume the same axisymmetric cavity with the 7-7-5 field gradient, cf. Fig. 2. From Table 1, the beam power for 1-A beam at the cavity exit is 2540 kW. Then the total RF input should be 3208 kW, and according to (3.1), the required cavity-waveguide coupling is $\beta_c$ = 4.80. From Eq. (3.3), the external quality factor for the pill-box model would be $Q_e$ = 556.

If four identical tapered RLWGs are used for RF input, as illustrated in Fig. 18, the RF input power per waveguide is 802 kW. Having four RLWGs instead of two already reduces $Q_e$ by a factor of 2, so we need only an extra factor 1.74 reduction, compared to the couplers used in Sect. 3, to obtain $Q_e$ = 556. This reduction (or equivalently, the coupling increase) can be achieved simply by increasing the diameter of the coupler iris holes to approximately 11-11.5 mm. A detailed design, of course, requires additional calculations. Most likely, the maximal power density on the coupler irises will be even lower than that in Sect. 3, since the iris holes are larger. This is due to the fact that the maximal power density on the coupler irises is defined mainly by the cavity fields, not by the RF power fed through the coupler. For larger holes, the magnetic field enhancement near the iris ends is smaller than for small ones.

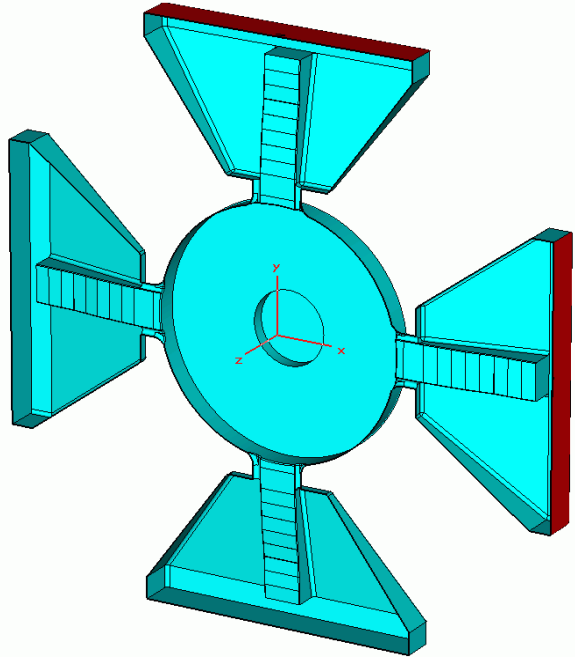

Fig. 18: MWS model with four tapered ridge-loaded waveguides: a cut-out view through the iris mid-plane ($z$ = 0) – cf. Fig. 8 with 2 RLWGs.

Running more than 800 kW RF power per RLWG will require two 1-MW CW klystrons and two RF windows per waveguide input. Due to the required control margin, the power available from one 1-MW klystron is typically around 750 kW, while a typical restriction for the available RF windows is 500 kW per window. In such a configuration, the photoinjector cavity will need 8 klystrons and 8 RF windows, with each klystron providing 400+ kW RF input power, unless the RF power from a smaller number of klystrons can be redistributed to feed four RF input waveguides.

As an illustration, we present some quick results obtained in a model having 4 RLWGs for RF power input, but using the same coupler irises (with 9.5-mm diameter holes) as in Sect. 3. In fact, this model was derived directly from the model in Fig. 8 by rotating the pillbox with 2 RLWGs by 90° around $z$-axis and then combining the resulting geometry with the original one in Fig. 8. The photoinjector cavity with such RF couplers would be matched for the electron beam current of 463 mA. This follows from the fact that with four RLWGs instead of two, the external quality factor $Q_e$ of the pill-box model is one-half of that in (3.3), i.e. $Q_e = 967$, and the coupling coefficient for the PI cavity is twice the value given by (3.1), i.e. $\beta_c = 2.76$. From Eq. (3.1), the beam power for the matched current is then 1176 kW, which corresponds to 0.463 A at energy 2.54 MeV. This beam current can be achieved with either 3.3 nC bunch charge at 140 MHz bunch repetition rate, or with 2.65 nC per bunch at 175 MHz. The total RF input is 1844 kW, or 461 kW per each of 4 RLWGs – obviously, the same RF input per waveguide as in the 2-RLWG model of Fig. 8. MWS time-domain calculations with a relatively crude mesh of 170K mesh points for 1/8 of the geometry confirm the reduction of $Q_e$ by a factor of 2. A snapshot of the surface currents at the match point is shown in Fig. 19, where the max power density near the iris ends corresponds to 93 W/cm$^2$. An eigensolver run gives 104 W/cm$^2$ with a similar mesh.

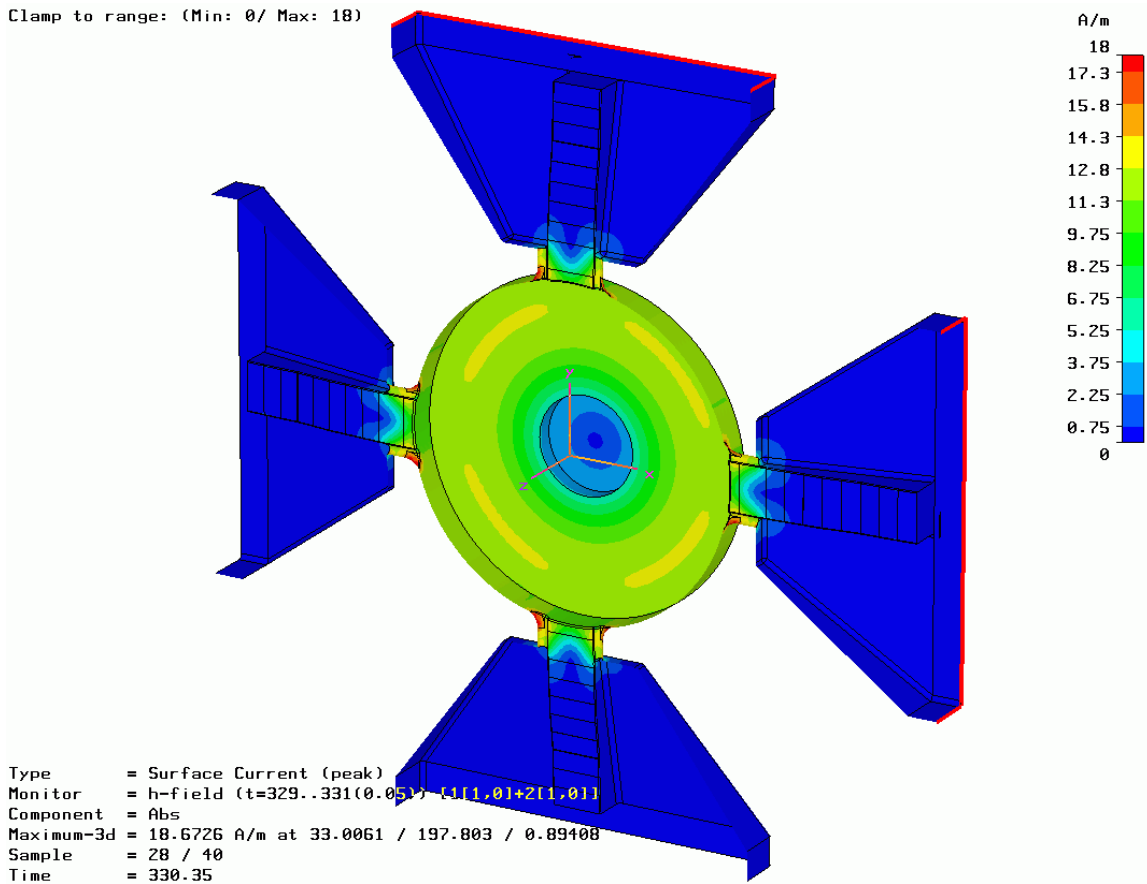

Fig. 19: Surface currents at the match point for MWS model with 4 RLWGs. Coupler irises have holes with diameter 9.5 mm, so that the PI cavity with this RF input is matched to 463-mA beam current.

The max density values are comparable to those in Tab. 2-3 with similar meshes, and will be somewhat higher with finer meshes. From a physical viewpoint, the maximal power density in this case should be very close to what was found for the model in Fig. 8, i.e. 120 W/cm$^2$. One should note that this photo-injector cavity can operate using 4 1-MW klystrons and 4 RF windows – one klystron and one window per RF input waveguide.

## 5. Summary.

A 100-mA CW operation of the normal-conducting RF cavity for the high-current CW photoinjector requires almost 1 MW of CW 700-MHz RF power. This RF power will be fed through two tapered ridge-loaded waveguides. The waveguide design was optimized taking into account manufacturing restrictions, which limit the taper length due to the brazing furnace dimensions.

The procedures used for designing the RF input couplers of the normal-conducting RF photoinjector are described in detail. The RF coupler design is based on the experience with the LEDA RFQ and SNS power couplers. The coupler-cavity system was modeled using a novel approach [8] with direct 3-D electromagnetic calculations using the CST MicroWave Studio in time domain. The results for the maximal power density are cross-checked using eigensolvers.

The RF coupler design is optimized using 3-D modeling to reduce the maximal power density on the coupler irises. This is achieved by increasing the radii of the end holes in the "dog-bone" shaped irises and by increasing the cavity wall thickness in the locations where the couplers are connected to the cavity. In addition, the outer edges of the iris holes are blended with a large blending radius inside the cavity. As the result, the magnetic field enhancement due to the coupler irises was kept at only 67% above the fields on the smooth cavity walls in the same cavity cross section.

For the 7-7-5 design of the photoinjector cavity, the maximal power density near the coupler irises is 120 $W/cm^2$, which is only about 15% higher than the maximal global power density in the smooth cavity without the RF couplers, 103 $W/cm^2$. These values are well below the max power density on the coupler irises in the LEDA RFQ, which was successfully operated in CW with 100 mA of the proton beam current.

## 6. References.